# Nanotwinned diamond synthesized from multi-core onion carbon

Qiang Tao, Xin Wei, Min Lian, Hongliang Wang, Xin Wang, Shushan Dong, Tian Cui and Pinwen Zhu*

State Key Laboratory of Superhard Materials, College of Physics, Jilin University, Changchun 130012, China

*Email: zhupw@jlu.edu.cn

**Abstract**

Nano-polycrystalline diamond (NPD) and nanotwinned diamond (NtD) were successfully synthesized in multi-anvil high pressure apparatus at high pressure and high temperature (HPHT) conditions using precursors of onion carbons. We found that the choices of distinct onion carbons with hollow or multi-cored microstructures lead to the synthesis of different diamond products of NPD or NtD ones. High quality NtD with an average twin size of 6.8 nm has been synthesized whose Vickers hardness reaches as high as 180 GPa as measured by indentation hardness experiment. The existence of stacking faults other than various defects in the onion carbon is found to be crucial to form twin boundary in the product. The origin of the extraordinarily high Vickers hardness in the NtD sample is attributable to the high concentration of twin boundary. Our work gave a direct support on the argument that pursuit of nanotwinned microstructure is an effective strategy to harden materials, in good coincidence with the well-known Hall-Petch effect.

**Introduction**

Superhard materials having high Vickers hardness (HV) more than 40 GPa, excellent abradability and chemical stability, are very important for industrial applications.[1-3] Diamond is the hardest material known with hardness ranging from 60 GPa to 120 GPa in Knoop hardness.[4-6] It is well-known that there is a major challenge to synthesize an alternative superhard material which is harder than diamond. However, grain boundary is known to impede the dislocation glide to resist the shape deformation, and thus is able to promote the material to have a higher hardness. According to the Hall-Petch effect, the mechanical properties of materials can be strengthened with decreasing grain size.[7,8] Therefore, fabrication of the nano-polycrystalline diamond (NPD) has been believed to be an effective method to enhance the mechanical properties of materials. For example, Koop hardness of NPD is 110~140 GPa when the grain size is 10~30 nm.[9] The HV of NPD is 147 GPa at the grain size of 13.4 nm. If the grain size reduces to 7 nm, the hardness was enhanced to be 167 GPa.[10] On the other hand, twin boundaries are considered to be the main factor for the enhanced hardness in NtD.[10-12] As a good example, it has been reported that the strength of nanotwinned copper has been improved five times better than coarse-grained copper [13-14]. Combining the Hall-Petch effect with twin boundaries, impressively, the Vickers hardness of high quality nanotwinned diamond (NtD) reaches remarkable 200 GPa with the twin size of 5 nm. Moreover, the fracture toughness and resistance of oxidation in NtD sample are also much improved.[15]

Both NPD and NtD are superior materials that can be categorized into ultra hard materials (HV≥80 GPa).[15-16] High pressure and high temperature (HPHT, P ≥10 GPa, T ≥1500℃) conditions are needed to synthesize both materials, while a proper choice on precursors is also neccessary.[17-19] As reported, NPD can be synthesized under HPHT by using polycrystalline graphite, black carbon, quasi-amorphous soot, glassy carbon balls or pyrolytic-graphite.[20] However, forming high quality NtD is strictly confined into the choice of onion carbon as precursor as reported by a pioneer work that introduces firstly the nanotwins into the microstructure of diamond.[15] In fact, not all of the onion carbons can directly transform into high quality NtD. For

example，the synthesized sample contains only partial twin boundaries with a large portion of NPD if the precursor is regular onion carbon.[10] By the witness of superior mechanical properties of NtD, it is essential to understand the transformation mechanism of onion carbon to NtD in an effort to fabricate high quality NtD.

In this work, different precursors of hollow onion carbon (HOC) and multi-core onion carbon (MOC) were prepared, and subsequently they were pressurized and heated in a large volume press for synthesis of NtD. The transformation mechanism to NtD was analyzed. Microstructure, morphology and hardness of the synthesized samples were characterized to optimize the synthetic conditions of high qulity NtD.

**Experimental**

HOC was synthesized by the starting materials of nano diamond (5 nm, 99.9 %), using the cubic anvil HPHT apparatus (SPD-6×600) under 1 GPa, 1100°C, 15 min. The MOC was fabricated by using black carbon powders through an impinging streams technology.[15] NPD and NtD was synthesized by large volume multi-anvil (Walker Type) under pressure of 20 GPa, temperature 2000°C~2300°C, holding time 2 min~30 min. The standard compress 10/4 sample assembly, a Re heater and a $LaCrO_3$ thermal insulator was used. Temperature was measured with type W-Re thermocouples, and pressure was estimated from previously obtained calibration curves for the multi-anvil apparatus. Recovered samples were 1.5 mm in diameter and 1 mm in height and polished for the further analysis. The Raman spectra were carried out to analysis the precursors and synthesized samples, and the samples were excited with 532 nm laser. A JEM-2200FS transmission electron microscope was used for TEM, HRTEM, and SAED observation. The Vickers microhardness measurements were performed by a Micro-Hardness Tester (HV-1000ZDT) and Hv was determined from $Hv=1854.4F/L^2$, where F is the applied load and L is the mean of the two diagonals of the indent in micrometers (μm). Atomic force microscope (AFM) using ScanAsyst mode in air (Dimension Icon, Veeco Instruments/Bruker) was used to further confirm the hardness value.[21]

**Results and Discussion**

NPD and NtD have been fabricated from different onion carbon at pressure of 20 GPa, temperature 2000°C~2300°C and holding time 2 min~30 min, respectively. The precursor of MOC transform to diamond under HPHT on the basis of Raman spectra, and without distinction with diamond synthesized by HOC in the mode of vibration (shown in Fig. 1). But the diamond substructure (synthesized by MOC) is entirely different with the NPD (synthesized by HOC). The microstructure for the as-synthesized sample was further analyzed by TEM and the results are shown in Fig. 2-4. It can be seen that the lamellar structure is shown in NPD synthesized from HOC (Fig. 2). The lamellar structure is also found in NPD synthesized from isotropic polycrystalline graphite rod.[22] Diamond layers in the lamellar structure are formed in the martensitic process from graphite via the hexagonal diamond phase.[23] Furthermore, partial twin boundary in NPD has been found (shown in Fig. S1). This result is in good agreement with previous reports.[10] It is worth noting that, in the TEM images of the sample synthesized from MOC (Fig. 3, Fig. 4), there are many multiple twins and stacking faults in the structure that indicate the synthesized sample is NtD. Most of the multiple twins are linear twins have the parallel twin boundary in the synthesized samples (NtD) (Fig. S2). But the twins present in NPD are mostly nonlinear twins that the twin boundaries are not parallel. And these twins in NPD are start at the surface of grain boundary and terminate at the same point in the grain.[20] The grain size range of NtD is from smaller than 1 nm to about 25 nm. The average of grain size is 6.8 nm (Fig.4 (b)).

The ultra hardness of NPD and NtD are always focused due to the application. The Vickers hardness was tested in synthesized NPD and NtD. Both the hardness of NtD and NPD are decrease with increasing of applied load (Fig. 5 (d)). The asymptotic hardness of NPD is 106 GPa consistent with single crystal hardness range of 60 GPa~120 GPa.[4-6] The grain boundary has not increase the hardness greatly. That is reasonable, the NPD has the lamellar substructure, which is thin in the thickness, but the size is reach micron scale. The grain boundary strengthen effect is results both by the thickness and size of lamellar. Therefore, the lamellar structure with big size results in the lower concentration of grain boundary that cannot contribute to enhance hardness. Moreover, the dislocation can slide along the grain boundary of lamellar structure, and results deformation. So, the hardness of lamellar NPD has lower hardness than the NPD (147 GPa) which has average size of 5 nm.[10]

The hardness of NtD is higher than NPD (Fig.5 (d)). The extremely high hardness of 180 GPa is uncovered under applied load of 4.9 N. In this case, the two diagonal lengths of an indentation are determined by optical microscope. This value is comparable with the previous result.[15] To confirm the extremely high hardness of NtD, the AFM, which can determine the diagonal lengths more accurate than optical microscope, was performed (Fig 5. (a)).[21] The diagonal line is 6.31 um in vertical and 6.62 um in horizontal under the indentation with the applied load 4.9 N, thus the hardness is 217 GPa reconfirmed by AFM (Fig. 5 (b) (c)). Obviously, the optically based measurement overestimates the diagonal length, thus providing a conservative measure of the Vickers hardness (underestimated by about 17.0%). This result is in good agreement with the previous report.[21] According to Hall-Petch relation, the higher hardness of NtD than that of NPD attribute to small average twin size of 6.8 nm which indicate high concentration twin boundary.[15,21,23-25] These results consistent with the indentation stress and strain relation that the twin boundaries in NtD promoting a large stress concentration which cause the bond transform to align in the hard shear direction.[26] With these results, the hardness is improved effectively by decrease the nano twin size, it is a appropriate method to fabricate ultra hard materials.

In order to understand the mechanism of different products formed from onion carbon under HPHT, it is very necessary to analyze the microtexture of onion carbon. Onion carbon for synthesizing NtD is fabricated by using black carbon powders through an impinging streams technology (Fig.6 (a)). Another for synthesizing NPD is synthesized at HPHT from nanodiamond (Fig.6 (b)). According to the HRTEM, the onion carbon from fabricated black carbon powders has the grain size of 20 nm~30 nm, moreover, inside one of the onion carbon, there are multi-core in it. These cores can form many dislocations and the stripe of MOC is not smooth but become waved, which results the stacking faults in the onion carbon. Both of dislocation and stacking faults can induce high defect concentration, which is benefit to diamond nucleation.[20] The average size of NtD (6.8 nm) is smaller than the MOC grain size about 20 nm~30 nm, but comparable with the core size in MOC. In the precursor of carbon black, the diamond nucleation can formed both at rim of carbon black and disordered core.[20] So, the diamond nucleation may occur both at the fringe and the core of MOC. Multiple cores in onion carbon results much of stacking faults, which may transfer to the twin boundary. In order to comparing with the multi-core in synthesizing NtD, HPHT was used to control the transformation of nano diamond to onion carbon. High pressure resistant the formation of the core in the onion carbon, create the ellipsoidal hollow onion carbon mixing much of open onion carbon (Fig.6 (b)). The grain size of HOC is 10 nm~15 nm. The character of HOC is no core and can achieve much of onion carbon grain boundary than MOC. But the stripe of HOC is smoother than MOC, which indicates lower concentration of stacking faults.

In order to further understand the character of the both onion carbon, Raman spectra was observed (Fig.7). The G mode at 1581 cm$^{-1}$ is corresponding to the first order scattering of the E$_{2g}$ mode. G mode is consistent with bond-stretching motion of pair of C that indicates *sp*$^2$ hybridization.[27-29] The D1 peak around 1355 is assigned to A$_{1g}$ mode indicate the integral vibration of six C-ring which cannot appear in perfect graphite structure.[27-29] The D2 mode around 1620 cm$^{-1}$ is apparent in HPHT onion carbon. Both the D1 and D2 mode is attribute to the disorder of the structure.[20] So the degree of defect concentration can express as:

$$R=D1/(G+D1+D2) \qquad (1)$$

G, D1, D2 is the area of each peak respectively.[20, 30] The ratio R of MOC and HOC is 54 % and 63 %, respectively. These results are consistent with HRTEM (Fig.1) that HOC contains more disorder structure due to the hollow and the more grain boundary (open onion carbon). And the hollow results the ellipsoidal or polygonal onion carbon to keep stable. Above all, the essential distinguish of MOC and HOC is that, 1. Both of MOC and HOC have high defect concentration, but HOC is higher; 2. The core in the onion carbon can cause much of stacking faults, which is absent in HOC.

Both the grain boundary and defect may important to diamond nucleation, however multi nucleation with same lattice orientation can grow fast under HPHT. The hollow reduce the mechanical properties of HOC, under high pressure, the ellipsoidal HOC will be closed to dense structure with no hollow, and the opened onion will reduce the curvature. Thus, the HOC precursor will be more like nano particle graphite with less of stacking faults under high pressure. Even though high defect concentration and high grain boundary density in HOC, which can fabricate multi of nucleation, the same lattice orientation of diamond nucleation can grow fast and is difficult to form twin boundary. So, although the high concentration of defect can form high density of nucleation, the grain will grow up under HPHT, and high concentration of stacking faults in the precursor is the essential reason to form high quality of NtD.

**Conclusion**

NPD and NtD were synthesized by HOC and MOC, respectively. The lamellar NPD has the hardness of 106 GPa. The NtD indicates high Vickers hardness about 180 GPa is attribute to high concentration of twin boundary. This result confirms that not all of onion carbon can synthesize NtD. Although there are many defects in both HOC and MOC, these defects in onion carbon are not the key to fabricate NtD, but the multi-cores which can results stacking faults in onion carbon is more important to synthesize NtD. The high defect concentration in HOC fabricates multi of nucleation with the same lattice orientation of diamond which grow fast under HPHT. So, it is difficult to form twin boundary but the lamellar microtexture. Further investigation to control the distribution of twin size in NtD is important to improve the integrity mechanical properties.

**Acknowledgements**

The authors acknowledge funding supports from the National Natural Science Foundation of China (No. 41572357). We acknowledge Dr Yongjun Bao for technical support with the Raman spectra measurements and Mrs Meihong Long for the technical support with AFM measurements.

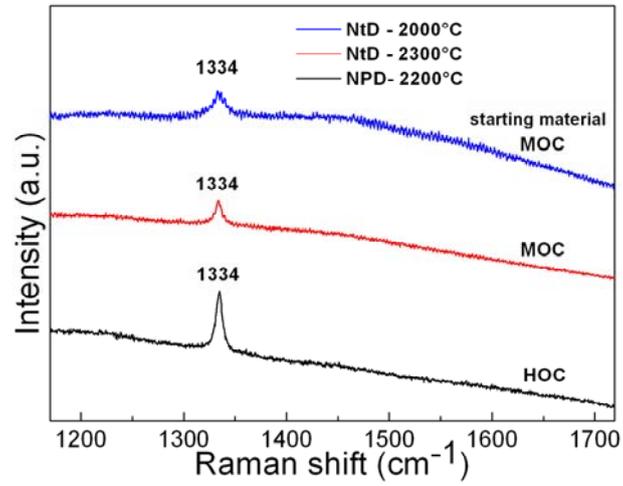

FIG. 1. Raman Spectra of synthesized diamond, from the top to bottom the starting material is MOC, MOC, and HOC respectively.

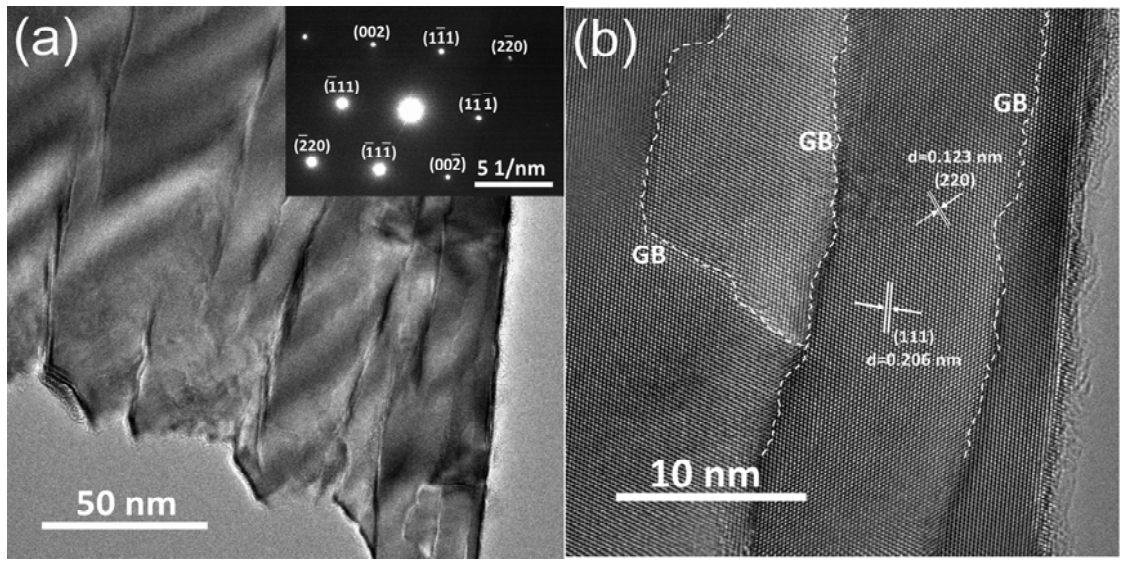

FIG. 2. (a) TEM of NPD synthesized by HOC, the insert pattern is SAED of NPD. (b) HRTEM of NPD.

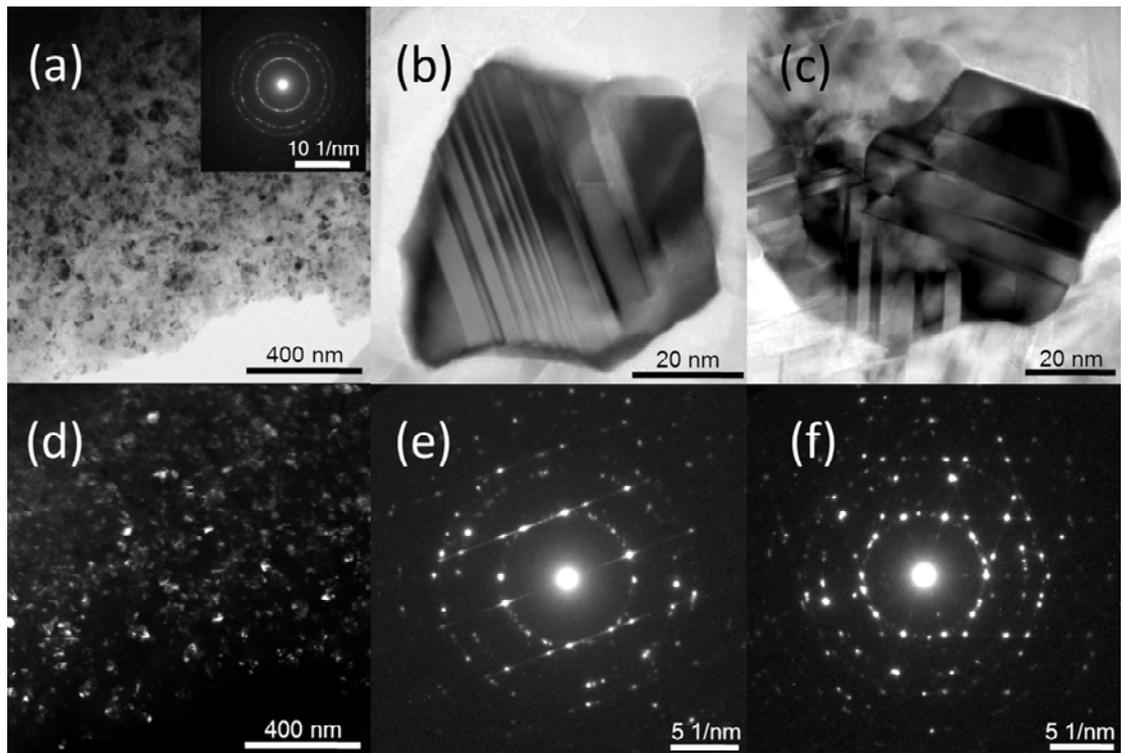

FIG. 3. (a) Bright Field TEM of NtD with the insert pattern of SAED which synthesized by MOC; (d) Dark Field TEM of NtD corresponding to (a); (b), (c) are the TEM of NtD and (e), (f) is SAED corresponding to (b), (c), respectively.

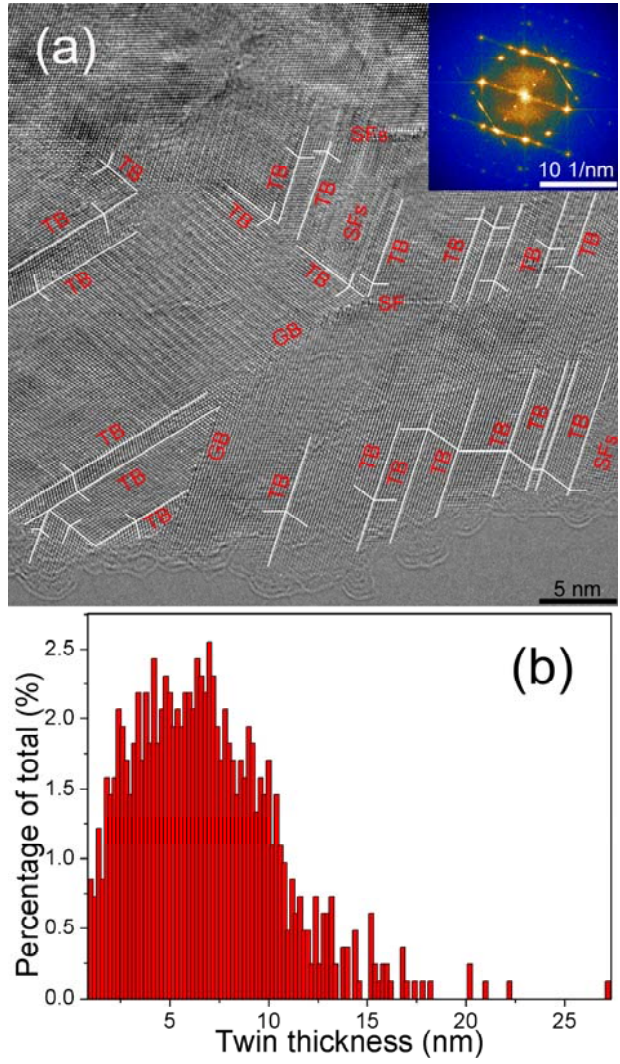

FIG. 4. (a) HRTEM of NtD synthesized by MOC (2000℃, 20 GPa), the insert pattern is SAED result. (b) the distribution of twin thickness.

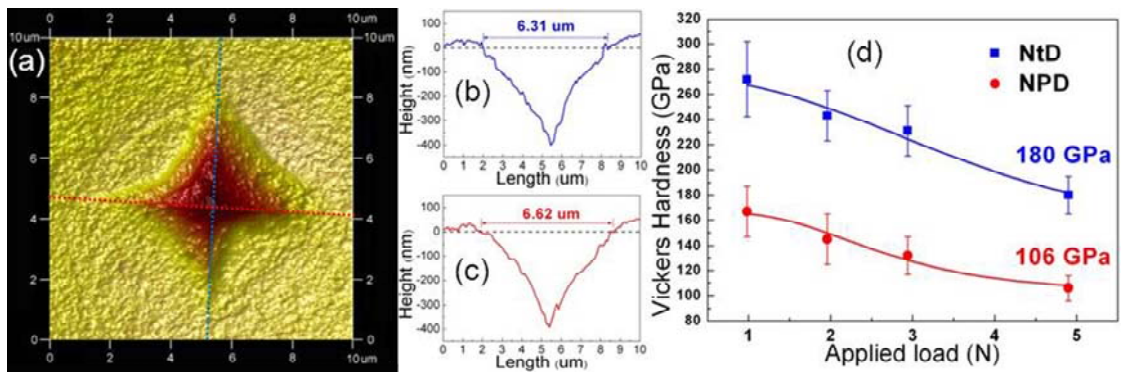

FIG. 5. (a) AFM of NtD with the applied load of 4.9 N. (b) the vertical diagonal line depth. (c) the horizontal diagonal line depth. (d) Vickers hardness of NtD and NPD.

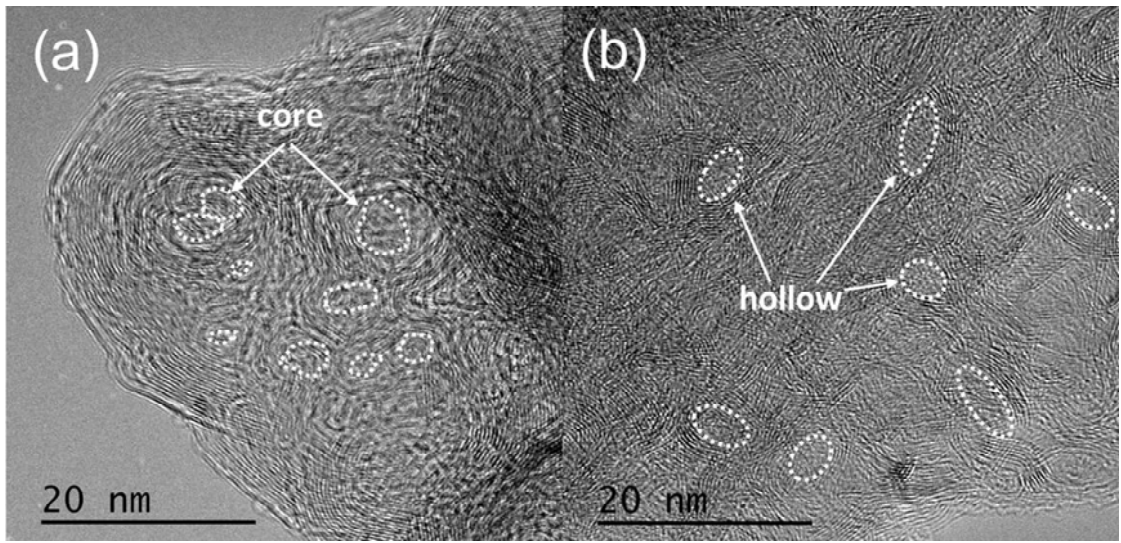

FIG. 6. The HRTEM of onion carbon, (a) multi-core onion carbon (MOC) fabricated by using black carbon powders through an impinging streams technology. (b) hollow-core onion carbon (HOC) synthesized by HPHT.

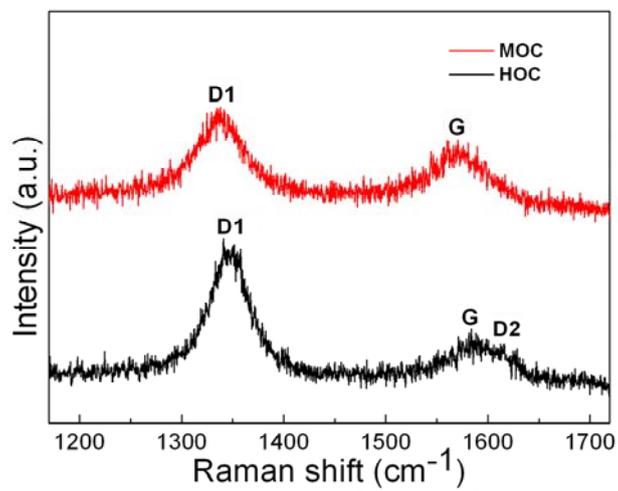

FIG. 7. Raman Spectra of MOC and HOC.

**Support information**

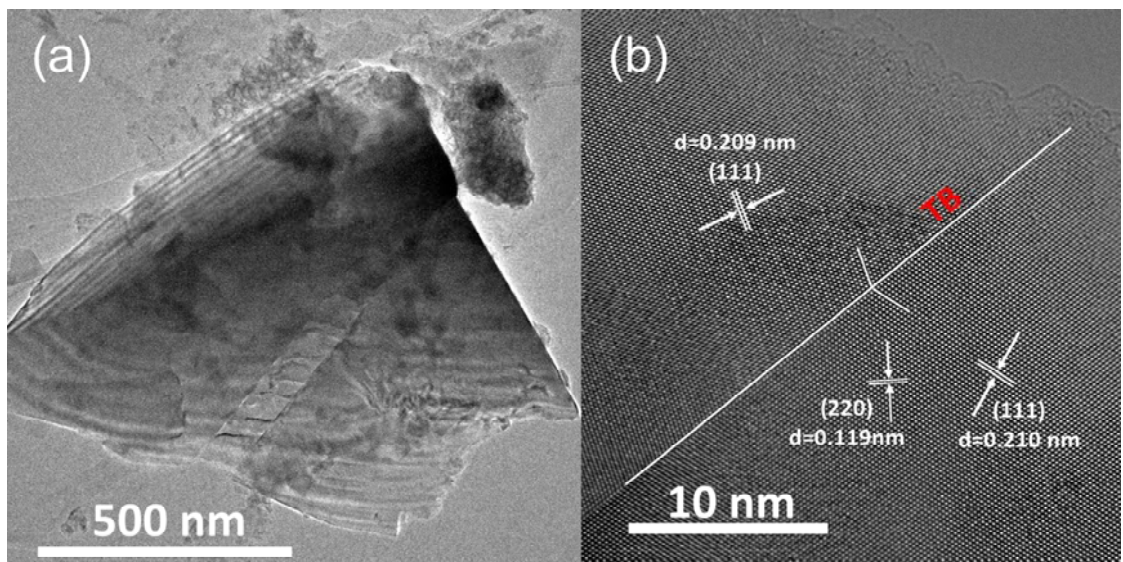

FIG. S1. (a) TEM of lamellar NPD. (b) HRTEM of NPD, partial twin boundary in NPD.

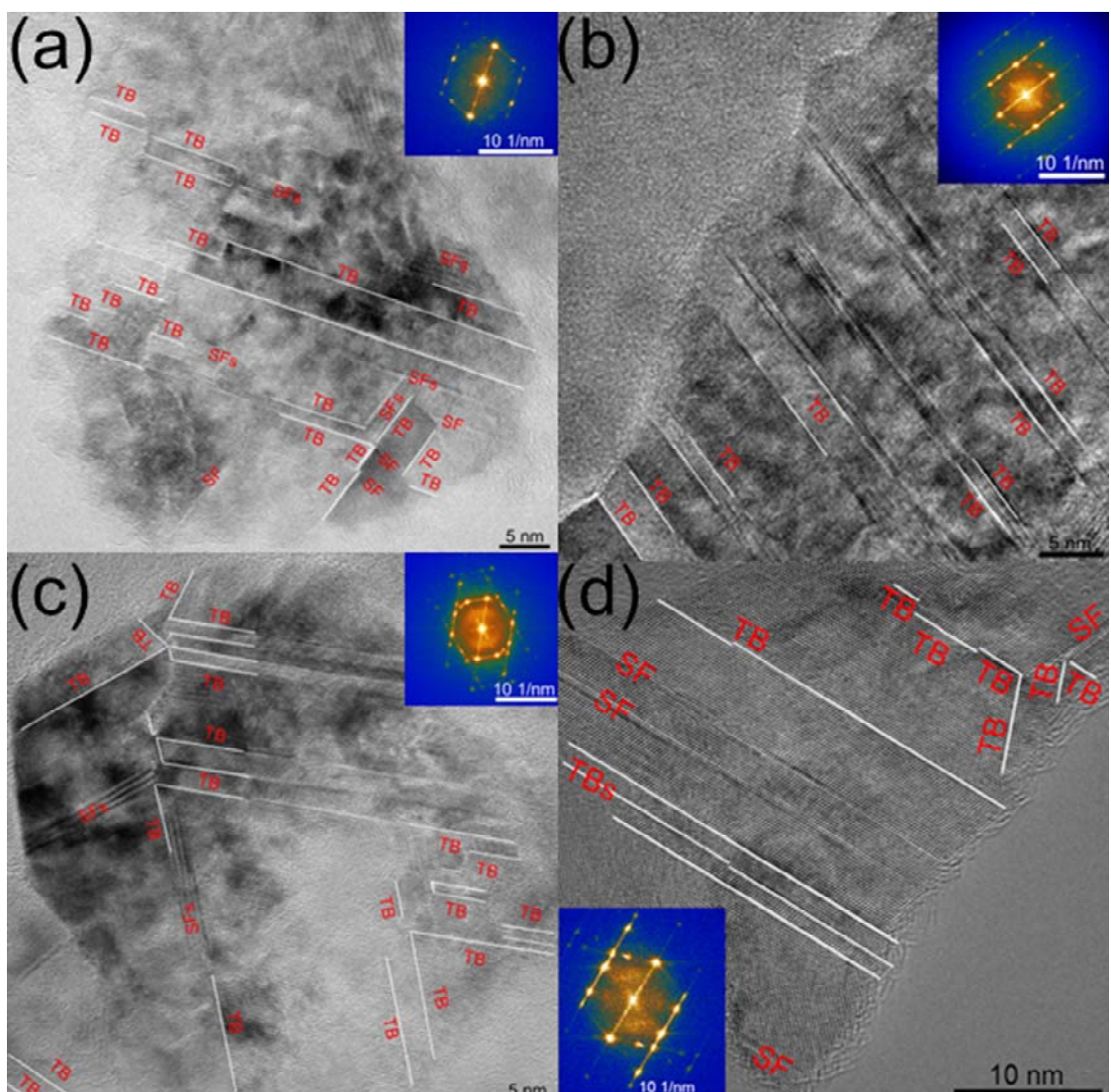

FIG. S2. (a), (b), (c), (d) are the different places of HRTEM in NtD, the insert patterns are the corresponding SAED results.